# Analysis of a Two-degree-of-freedom Beam for Rotational Piezoelectric Energy Harvesting

Xiang-Yu Li, I-Chie Huang, Wei-Jiun Su*

Department of Mechanical Engineering, National Taiwan University, No. 1, Sec. 4, Roosevelt Rd., Taipei, 10617, Taiwan

**Abstract**

This study introduces a two-degree-of-freedom piezoelectric energy harvester designed to harness rotational motion as an energy source. The harvester is built using a cut-out beam, which enables the first two resonant frequencies to be closely located in the low-frequency range. A distributed continuous model is developed and validated with experimental results. As the beam undergoes significant displacement due to rotational excitations, the geometric nonlinearity arising from longitudinal displacement is considered in the model to enhance its accuracy. It is observed that as the rotating speed increases, the increased centrifugal force causes the first resonant frequency to rise while the second resonant frequency decreases. The rotation-specific mode veering and the interchange of the first two modes are discussed. This study explores the potential to expand the bandwidth of the harvester using two types of nonlinear external force, namely mechanical stoppers and magnetic force. The results indicate that the proposed harvester can broaden the bandwidth of the first and second resonant frequencies. This research addresses the gap of combining multimodal and nonlinear force methods in rotational piezoelectric energy harvesting.

## 1. Introduction

Over the past few decades, with the rapid development of wireless autonomous sensor systems, batter power suffers from regular replacement and environment contamination. Among various energy harvesting techniques, vibration-based piezoelectric energy harvesters (PEHs) offer several advantages over other solutions, including downsizing capabilities and high power density [1]. The cantilevered beam is a widely used structure for piezoelectric energy harvesting due to its compactness and simplicity of mechanism. Erturk and Inman et al. [2, 3] proposed a theoretical model of a cantilevered PEH under base excitations and validated it with experimental results. This model, however, underestimates deflection and overestimates natural frequencies due to the neglect of transverse shear deformation and rotary inertia [4]. Compared to base excitation, rotational energy harvesting has the potential applications in various fields, including vehicle wheels, machine tools, and wind turbines. Yigit et al. [5] derived fully coupled nonlinear equations for a rotating cantilevered beam and investigated the impact of coupling terms on its dynamic responses. Gu and Livermore et al. [6] verified the effect of centrifugal force in a rotating environment on a cantilevered PEH, finding that its resonant frequency closely matches the rotation frequency over a wide range from 4 to 16.2 Hz. Khameneifar et al. [7, 8] developed a multimode mathematical model for the coupled electro-mechanical rotational system. Hsu et al. [9] employed finite element analysis to study a rotating cantilevered beam with an end mass and validated their results with experimental data. Mei et al. [10] uncovered the underlying mechanism of the centrifugal force acting on the PEH and demonstrated its impact on dynamic performances. Su et al. [11] explored three orientations, namely inward, outward, and tilted configurations, for a horizontal rotational PEH. According to the findings, adjusting the installation distance and tilt angle can match the system's resonant frequency with the excitation frequency.

Linear PEHs only work efficiently when the excitation frequency is close to the natural frequencies, limiting the harvesting bandwidth. To address this challenge, recent studies on rotational PEHs have focused on broadening the bandwidth of rotational energy harvesting through various methods, including mechanical structure improvement [12-15], external circuitry [16], nonlinear methods [17-32], and multi-modal techniques [33-35]. Nonlinear methods can enhance rotational energy harvesting by leveraging the system's inherent characteristics, such as the hardening/softening effect in rotational systems. Mei et al. [17] propose an inverse PEH that exhibits a centrifugal softening effect, significantly enhancing the output voltage over the speed range of 75-120 rpm (1.2-2 Hz). Additionally, nonlinear forces can be introduced externally, such as magnetic forces [18-26] and impact forces [28-32].

Magnetic force is commonly used to introduce nonlinearity into rotational energy harvesters. Kan et al. [18] utilized repulsive magnetic force to axially excite a cantilevered PEH in rotary motion for energy harvesting. Recent research on rotational energy harvesting with magnetic forces has increasingly focused on multi-stable systems by constructing potential energy fields with magnet arrays [19]. The low potential barrier of the multi-stable energy harvester facilitates achieving inter-well oscillations through time-varying potential wells [20]. Mono-stable [21] and bi-stable [22-24] were initially predominant but investigations have rapidly expanded to include more stable states. Mei et al. investigated the performance of a tri-stable PEH [25] and a quad-stable PEH [26] in rotational motion, combining the advantages of lower potential barriers and time-varying potential wells. Impact force is another commonly used nonlinear method that significantly contributes to widening the bandwidth. Rui et al. [28] examined a rotational energy harvester with limiters and discovered that stiffness has little impact on the limiting effect. Fang et al. [29, 30] analyzed a rotational impact energy harvester composed of a centrifugal-softening driving beam. They found that the centrifugal softening effect can amplify the relative motion and increase the impact force, thereby improving output power at low rotational frequencies. Machado et al. [31] proposed a low frequency rotational PEH that can operate out of resonance by incorporating a flexible stop.

Multi-modal technology has been widely studied as a promising method for wideband energy harvesting. Wu et al. [33] design a beam with a cut-out section and an inverted secondary beam inside, achieving two closely located resonant frequencies at low frequency, which inspired this research. Yu et al. [34] propose a novel PEH with rotating-degree-of-freedom to efficiently achieve tri-modal vibration energy harvesting in low-frequency environments, particularly at its first resonant frequency, down to 12.65 Hz. Raja et al. [35] utilize a multi-mode structure comprising a reversed exponentially tapered beam and six branched beams attached with a flange to the free end of the primary beam.

In this study, a two-degree-of-freedom (2-DOF) rotational PEH that incorporates the multi-modal method and nonlinearities is proposed. This PEH aims to have two closely-located resonant frequencies to achieve broadband harvesting under rotational excitation. The centrifugal force resulting from rotational motion is expected to bring the first two resonant frequencies even closer. The model incorporates geometric nonlinearity due to large deflection. Two types of nonlinear force, impact force caused by stoppers and nonlinear magnetic force, are introduced to examine their respective influences on the performance of the PEH.

## 2. Design and modeling

As shown in Fig. 1(a), the proposed design is composed of a cut-out beam, a piezoelectric layer, and two tip masses. The cut-out beam is divided into four segments, with the main beam fixed on the rotational base. The piezoelectric layer is attached to the second segment of the main beam to facilitate comparison of tests on the same prototype with different lengths of the first segment by clamping at different locations. Two tip masses ($Mt_3$/$Mt_4$) are mounted at the free end of the main beam and auxiliary beam. It is noted that the model is capable of accommodating four tip masses attached to the end of the four segments, but the first two tip masses ($Mt_1$/$Mt_2$) are set as zero to form a 2-DOF system. In the subsequent discussion, the main beam refers to the range covering $L_1$ to $L_3$, while the auxiliary beam covers $L_4$. The local coordinates of the PEH are depicted in Fig. 1(b), where $x_i$ denotes the longitudinal position of the beam along the neutral axis.

Furthermore, this study aims to investigate the effects of stoppers and magnetic force on the performance of the PEH. The schematics of the PEHs equipped with stoppers and magnets under rotational excitations are shown in Fig. 1(c-f). The stopper is a cantilever beam with a tip mass attached to it for collision. The fixed magnet is mounted directly below the magnet of the cut-out beam. It can be observed that the main beam is oriented outward from the rotating center, while the auxiliary beam is oriented inward toward the rotating center. These two distinct orientations will result in differential effects of the centrifugal force on the beams, which will be thoroughly examined in this study.

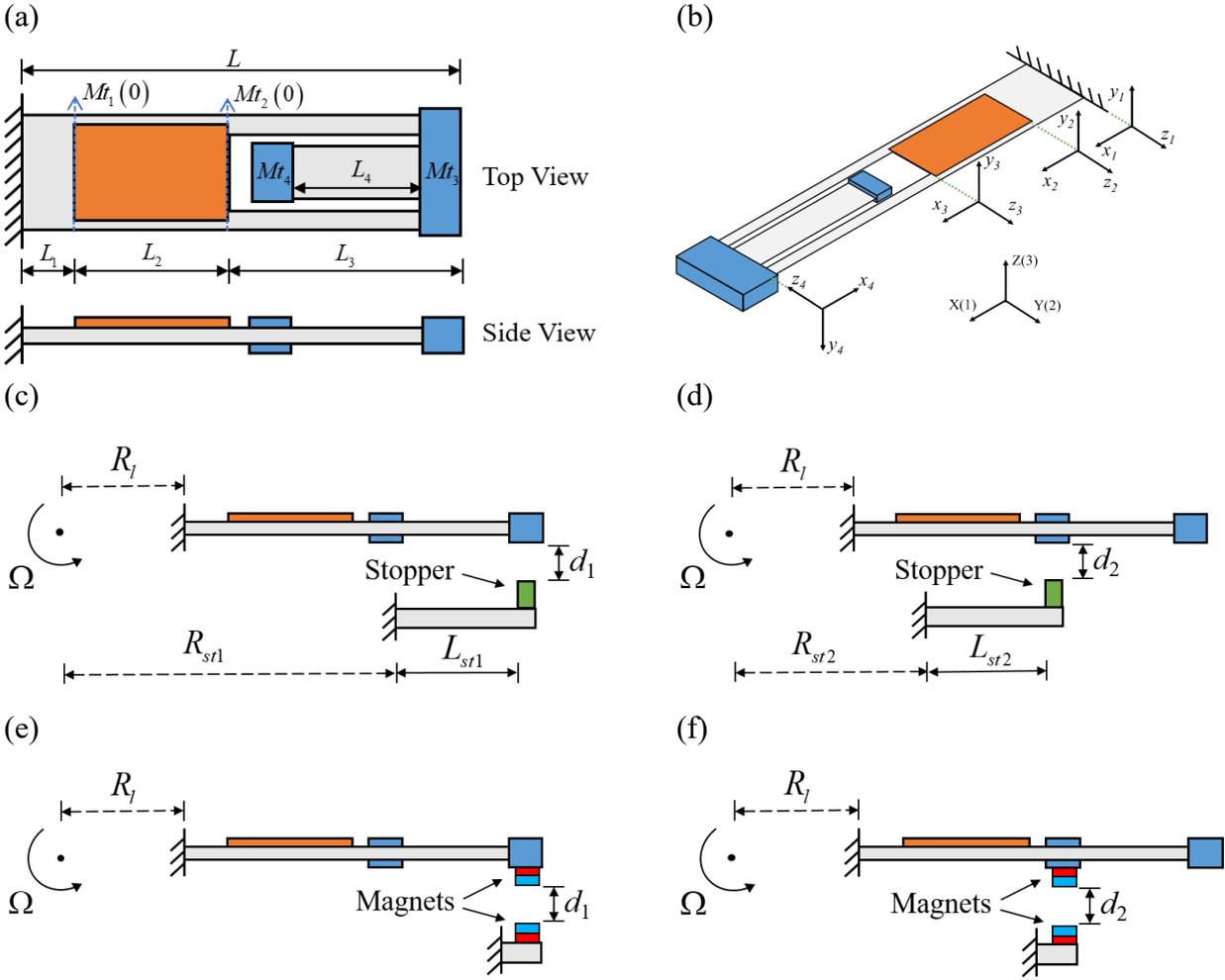

Fig. 1 Schematics of the proposed PEH: (a) top and side views (b) local coordinates of the PEH (c) the PEH with a stopper on the main beam (d) the PEH with a stopper on the auxiliary beam (e) the PEH with magnets on the main beam (f) the PEH with magnets on the auxiliary beam.

**2.1. PEH with Geometric Nonlinearity**

When subjected to rotational excitations, the transverse displacement of the PEH is significant due to the substantial excitation force caused by gravity. Therefore, this paper presents a theoretical model that incorporates the longitudinal displacement of the beam. Consider an infinitesimal beam element as shown in Fig. 2. Assuming that the beam is in-extensional, the longitudinal displacement can be expressed using geometrical relationships [36] as follows:

$$u_i(x_i,t) = \int_0^{x_i} \left(\sqrt{1-w_i'^2(s,t)} - 1\right)ds \approx -\frac{1}{2}\int_0^{x_i} w_i'^2(s,t)ds \tag{1}$$

where $w$ is the transverse displacement, $u$ is the longitudinal displacement, $ds$ is the length of the infinitesimal element, and the subscript $i$ indicates the segment number. It is noted that the X and Z indicate the longitudinal and transverse directions of the undeformed element, respectively.

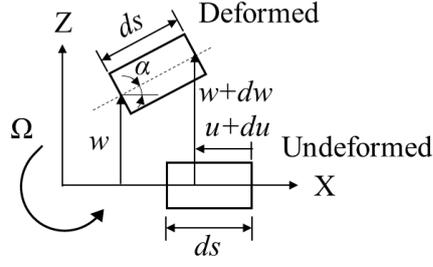

Fig. 2 Schematics of an infinitesimal beam element

The force equilibrium of an infinitesimal element can be referred to Appendix A. The moment equilibrium of the beam element can be expressed as:

$$dM_i - P_i \times dw_i + Q_i \times dx_i = m_i I_i dx_i \ddot{\alpha} \tag{2}$$

where $Q_i$ is the shear force, $I_i$ is the area moment of inertia, $M_i$ is the bending moment, $m_i$ is the mass per unit length of the beam, and $\alpha$ is the angle of the deformed beam element and can be expressed as:

$$\alpha = \tan^{-1}\left(\frac{w'}{1+u'}\right) \approx w' \tag{3}$$

The centrifugal force $P_i$ can be written as:

$$P_i = \left[ m_i \int_{x_i}^{L_i}(x_i + R_i)dx_i + Mt_i(L_i + R_i) \right]\Omega^2 \tag{4}$$

where $\Omega$ is the rotational speed, $R$ is the distance between the rotating center to the origin of the beam segment, and $Mt_i$ is tip mass. The shear force $Q_i$ can then be expressed as:

$$Q_i = -YI_i\left[ w_i'''(x_i,t) + \frac{1}{2}w_i'''(x_i,t)w_i'^2(x_i,t) + w_i'(x_i,t)w_i''^2(x_i,t) \right] + P_i w_i'(x_i,t) + m_i I_i \ddot{w}_i'(x_i,t) \tag{5}$$

where $YI_i$ is the bending stiffness.

Integrate the Z-direction force equilibrium equation (eqn.(A1)) from $x_i$ to $L_i$ and incorporate it into the Z-direction force equilibrium equation at the tip mass (eqn.(B1)). The relationship between $F_i$ and $Q_i$ can then be expressed as:

$$F_i(x_i,t) = \begin{cases} \dfrac{1}{\cos(\alpha_i(x_i,t))}\left\{ \begin{array}{l} Q_i \sin(\alpha_i(x_i,t)) + [fc_i(x_i) + fc_{i+1}(0)]\Omega^2 \\ -[fg_i(x_i) - fg_{i+1}(0)]\cos\Omega t + fn_i(x_i,t) + fn_{i+1}(0,t) \end{array} \right\} & i = 1,2,3 \\ \dfrac{1}{\cos(\alpha_4(x_4,t))}\left[ Q_4 \sin(\alpha_4(x_4,t)) + fc_4(x_4)\Omega^2 + fg_4(x_4)\cos\Omega t + fn_4(x_4,t) \right] & i = 4 \end{cases} \tag{6}$$

where $fc_i(x_i)\Omega^2$ is the centrifugal force coefficient, $fg_i(x_i)\cos\Omega t$ is the axial component of gravity coefficient, and $fn_i(x_i,t)$ is the axial component of inertia coefficient. $\sin(\alpha)$, $\cos(\alpha)$ and the aforementioned terms can be expressed as:

$$\sin(\alpha) = w' \tag{7a}$$

$$\cos(\alpha) = 1 + u' \approx 1 - \frac{1}{2}w'^2 \tag{7b}$$

$$fc_i(x_i) = \begin{cases} m_i \int_s^{L_i} (x_i + R_i)dx_i + Mt_i(L_i + R_i) & i=1,2,3 \\ -m_4 \int_s^{L_4} (x_4 + R_4)dx_4 + Mt_4(L_4 + R_4) & i=4 \end{cases} \tag{8a}$$

$$fg_i(x_i) = \begin{cases} \left[(L_i - x_i)m_i + Mt_i\right]g & i=1,2,3 \\ -\left[(L_4 - x_4)m_4 + Mt_4\right]g & i=4 \end{cases} \tag{8b}$$

$$fn_i(x_i,t) = \begin{cases} m_i \int_s^{L_i} \left[\frac{\partial^2 u_i(x_i,t)}{\partial t^2} - \Omega^2 u_i(x_i,t)\right]ds + Mt_i\left[\frac{\partial^2 u_i(L_i,t)}{\partial t^2} - \Omega^2 u_i(L_i,t)\right] & i=1,2,3 \\ -m_4 \int_s^{L_4} \left[\frac{\partial^2 u_4(x_4,t)}{\partial t^2} - \Omega^2 u_4(x_4,t)\right]ds - Mt_4\left[\frac{\partial^2 u_4(L_4,t)}{\partial t^2} - \Omega^2 u_4(L_4,t)\right] & i=4 \end{cases} \tag{8c}$$

Substitute eqns. (5) to (6) into the force equilibrium equation (eqn.(B2)), the forced vibration equation of the cut-out beam can be obtained:

$$\begin{cases} YI_i\left\{w_i''''(x_i,t) + \left[w_i'(x_i,t)\left(w_i'(x_i,t)w_i''(x_i,t)\right)'\right]'\right\} + m_i\left[\frac{\partial^2 w_i(x_i,t)}{\partial t^2} - \Omega^2 w_i(x_i,t)\right] & i=1,2,3 \\ -AF_i w_i''(x_i,t) = m_i g \sin\Omega t \\ YI_4\left\{w_4''''(x_4,t) + \left[w_4'(x_4,t)\left(w_4'(x_4,t)w_4''(x_4,t)\right)'\right]'\right\} + m_4\left[\frac{\partial^2 w_4(x_4,t)}{\partial t^2} - \Omega^2 w_4(x_4,t)\right] \\ -AF_4 w_4''(x_4,t) = -m_4 g \sin\Omega t \end{cases} \tag{9}$$

where the axial force $AF_i$ can be expressed as:

$$AF_i(x_i,t) = \begin{cases} \left[fc_i(x_i) + fc_{i+1}(0)\right]\Omega^2 - \left[fg_i(x_i) - fg_{i+1}(0)\right]\cos\Omega t + \left[fn_i(x_i,t) + fn_{i+1}(0,t)\right] & i=1,2,3 \\ fc_4(x_4)\Omega^2 + fg_4(x_4)\cos\Omega t + fn_4(x_4,t) & i=4 \end{cases} \tag{10}$$

### 2.2. Modal Analysis

Neglecting the damping, the external force, and the nonlinear terms, the undamped free vibration of the beam can be rewritten as:

$$\begin{cases} YI_i w_i''''(x_i,t) - \left[fc_i(x_i) + fc_{i+1}(0)\right]\Omega^2 w_i''(x_i,t) + m_i\left(\frac{\partial^2 w_i(x_i,t)}{\partial t^2} - \Omega^2 w_i'(x_i,t)\right) = 0 & i=1,2,3 \\ YI_4 w_4''''(x_4,t) - fc_4(x_4)\Omega^2 w_4''(x_4,t) + m_4\left(\frac{\partial^2 w_4(x_4,t)}{\partial t^2} - \Omega^2 w_4'(x_4,t)\right) = 0 & i=4 \end{cases} \tag{11}$$

where the transverse displacement $w_i$ can be expressed as:

$$w_i(x_i,t) = \sum_{j=1}^{n} \phi_{ij}(x_i)\eta_j(t) \tag{12}$$

where $\phi_{ij}$ is the mode shape function, $\eta_j$ is the temporal function, and the subscript $j$ represents the mode number. As the centrifugal force term $fc_i(x_i)$ in eqn. (11) is non-constant, it needs to be converted into a constant for modal analysis. In this study, this term is replaced by its RMS value based on eqn.(13):

$$\overline{fc_i} = \begin{cases} \sqrt{\dfrac{\int_0^{L_i}\left[fc_i(x_i)+fc_{i+1}(0)\right]^2 dx_i}{L_i}} & i=1,2,3 \\ \sqrt{\dfrac{\int_0^{L_4} fc_4(x_4)^2 dx_4}{L_4}} & i=4 \end{cases} \tag{13}$$

The model shape function can be expressed as:

$$\phi_{ij}(x_i) = A_{ij}\cosh(a_{ij}x_i) + B_{ij}\sinh(a_{ij}x_i) + C_{ij}\cos(b_{ij}x_i) + D_{ij}\sin(b_{ij}x_i) \tag{14}$$

where $A_{ij}$, $B_{ij}$, $C_{ij}$, and $D_{ij}$ will be determined by the boundary and continuous conditions. The boundary conditions of the rotational PEH are:

$$w_1(0,t) = 0 \tag{15a}$$

$$w_1'(x_1,t)\big|_{x_1=0} = 0 \tag{15b}$$

$$w_4''(x_4,t)\big|_{x_4=L_4} = 0 \tag{15c}$$

$$YI_4\left[w_4'''(x_4,t)w_4'^2(x_4,t) + \left(w_4''(x_4,t)\right)^2 w_4'(x_4,t) + w_4'''(x_4,t)\right] - AF_4 w_4'(x_4,t) - Mt_4 g \sin\Omega t$$
$$-Mt_4\left[\dfrac{\partial^2 w_4(x_4,t)}{\partial t^2} - \Omega^2 w_4(x_4,t)\right] = 0 \tag{15d}$$

The continuous conditions of the rotational PEH are:

$$w_1(L_1,t) = w_2(0,t) \tag{16a}$$

$$w_1'(x_1,t)\big|_{x_1=L_1} = w_2'(x_2,t)\big|_{x_2=0} \tag{16b}$$

$$YI_1 w_1''(x_1,t)\big|_{x_1=L_1} = YI_2 w_2''(x_2,t)\big|_{x_2=0} \tag{16c}$$

$$YI_1\left[w_1'''(x_1,t)w_1'^2(x_1,t) + w_1''^2(x_1,t)w_1'(x_1,t) + w_1'''(x_1,t)\right] - AF_1(x_1)w_1'(x_1,t) + Mt_1 g \sin\Omega t$$
$$-Mt_1\left[\dfrac{\partial^2 w_1(x_1,t)}{\partial t^2} - \Omega^2 w_1(x_1,t)\right]\bigg|_{x_1=L_1} = YI_2 w_2''(x_2,t) - AF_2(x_2)w_2'(x_2,t)\big|_{x_2=0} \tag{16d}$$

$$w_2(L_2,t) = w_3(0,t) \tag{16e}$$

$$w_2'(x_2,t)\big|_{x_2=L_2} = w_3'(x_3,t)\big|_{x_3=0} \tag{16f}$$

$$YI_2 w_2''(x_2,t)\Big|_{x_2=L_2} = YI_3 w_3''(x_3,t)\Big|_{x_3=0} \tag{16g}$$

$$YI_2\left[w_2'''(x_2,t)w_2'^2(x_2,t)+w_2''^2(x_2,t)w_2'(x_2,t)+w_2'''(x_2,t)\right]-AF_2(x_2)w_2'(x_2,t)+Mt_2 g\sin\Omega t$$
$$-Mt_2\left[\frac{\partial^2 w_2(x_2,t)}{\partial t^2}-\Omega^2 w_2(x_2,t)\right]\Bigg|_{x_2=L_2} = -YI_3 w_3''(x_3,t)+AF_3(x_3)w_3'(x_3,t)\Big|_{x_3=0} \tag{16h}$$

$$w_3(L_3,t) = -w_4(0,t) \tag{16i}$$

$$w_3'(x_3,t)\Big|_{x_3=L_3} = w_4'(x_4,t)_{x_4=0} \tag{16j}$$

$$YI_3 w_3''(x_3,t)\Big|_{x_3=L_3} + I_{M_3}\frac{\partial^2}{\partial t^2}w_3'(x_3,t)\Big|_{x_3=L_3} = YI_4 w_4''(x_4,t)\Big|_{x_4=0} \tag{16k}$$

$$YI_3\left[w_3'''(x_3,t)w_3'^2(x_3,t)+w_3''^2(x_3,t)w_3'(x_3,t)+w_3'''(x_3,t)\right]-AF_3(x_3)w_3'(x_3,t)+Mt_3 g\sin\Omega t$$
$$-Mt_3\left[\frac{\partial^2 w_3(x_3,t)}{\partial t^2}-\Omega^2 w_3(x_3,t)\right]\Bigg|_{x_3=L_3} = -YI_4 w_4''(x_4,t)+AF_4(x_4)w_4'(x_4,t)\Big|_{x_4=0} \tag{16l}$$

where the rotational inertia of the $i^{th}$ beam segment $I_{Mi}$ can be represented as:

$$I_{M_i} = Mt_i I_i \ddot{\alpha}_i(x_i,t)\Big|_{x_i=L_i} \approx Mt_i I_i \frac{\partial^2}{\partial t^2}w_i'(x_i,t)\Big|_{x_i=L_i} \tag{17}$$

Eqns.(15a)-(16l) can be rearranged into a matrix form as follows:

$$M_{16\times 16} \cdot \begin{bmatrix} A_{ij} \\ \vdots \\ D_{ij} \end{bmatrix} = \begin{bmatrix} 0 \\ \vdots \\ 0 \end{bmatrix}, \tag{18}$$

The natural frequencies can be determined by setting the determinant of $M_{16\times 16}$ equal to zero. The mode shape function is then mass-normalized according to the orthogonal equation:

$$\begin{cases} \sum_{i=1}^{4}\left\{\int_0^{L_i}\phi_{ik}(x_i)m_i\phi_{ij}(x_i)dx_i + \phi_{ik}(x_i)Mt_i\phi_{ij}(x_i)\Big|_{x_i=L_i} + \left[\phi_{ik}'(x_i)I_{Mt_i}\phi_{ij}'(x_i)dx_i\right]\right\} = \delta_{jk} \\ \sum_{i=1}^{3}\left\{\int_0^{L_i}\phi_{ik}'(x_i)\left[fc_i(x_i)+fc_{i+1}(0)\right]\Omega^2\phi_{ij}'(x_i)dx_i\right\} + \int_0^{L_4}\phi_{4k}'(x_4)fc_4(x_4)\Omega^2\phi_{4j}'(x_4)dx_4 \\ +\sum_{i=1}^{4}\left\{\int_0^{L_i}\phi_{ik}''(x_i)YI_i\phi_{ij}''(x_i)dx_i + \left[\phi_{ik}'(x_i)I_{Mt_i}\phi_{ij}''(x_i)dx_i\right] \\ -\Omega^2\left[\int_0^{L_i}\phi_{ik}(x_i)m_i\phi_{ij}(x_i)dx_i + \phi_{ik}(x_i)Mt_i\phi_{ij}(x_i)\Big|_{x_i=L_i}\right]\right\} \end{cases} = \omega_j^2\delta_{jk} \quad \delta_{jk}=\begin{cases}0 & if\ j=k \\ 1 & if\ j\neq k\end{cases} \tag{19}$$

The undamped equation of motion for the PEH is then derived as:

$$\ddot{\eta}_j(t) + \omega_j^2 \left(1 - \frac{Kg_{jk}\cos\Omega t}{Kb_{jk} + (Ks_{jk} - Mb_{jk})\Omega^2}\right)\eta_j(t) + \left(Rs_{jk} - \frac{1}{2}\Omega^2 Kn_{jk}\right)\eta_j^3(t)$$
$$+ Kn_{jk}\left[\eta_j(t)\dot{\eta}_j^2(t) + \eta_j^2(t)\ddot{\eta}_j(t)\right] = F_j \sin\Omega t \tag{20}$$

where $Mb_{jk}$, $Kb_{jk}$, $Ks_{jk}$, $Kg_{jk}$, $Rs_{jk}$, $Kn_{jk}$, and $F_j$ represent the terms of quality, rigidity, stress hardening, gravitational force, rigidity nonlinearity, inertia nonlinearity and external force. The details of these terms can be found in Appendix C. The first two modes are investigated in this study. Assuming that the first two modes are perfectly orthogonal, only the principal diagonal terms of these matrices are nonzero and we can transform those terms from $2\times2$ matrices to $2\times1$ vectors, taking $Kb_j$ as an example:

$$Kb_{2\times2} = \begin{bmatrix} Kb_{11} & Kb_{12} \\ Kb_{21} & Kb_{22} \end{bmatrix} \approx \begin{bmatrix} Kb_{11} & 0 \\ 0 & Kb_{22} \end{bmatrix} \Rightarrow Kb_{2\times1} = \begin{bmatrix} Kb_1 \\ Kb_2 \end{bmatrix} = \begin{bmatrix} Kb_{11} \\ Kb_{22} \end{bmatrix} \tag{21}$$

Finally, considering damping, excitation force, and electromechanical coupling, the governing equation of the PEH can be rewritten as:

$$\ddot{\eta}_j(t) + 2\zeta_j\omega_j\dot{\eta}_j(t) + (\omega_j^2 - Kg_j\cos\Omega t)\eta_j(t) + \left(Rs_j - \frac{1}{2}\Omega^2 Kn_j\right)\eta_j^3(t)$$
$$+ Kn_j\left[\eta_j(t)\dot{\eta}_j^2(t) + \eta_j^2(t)\ddot{\eta}_j(t)\right] + \theta_j v(t) = F_j \sin\Omega t \qquad j=1,2 \tag{22}$$

where $\zeta_j$ is the mechanical damping ratio and $F_j$ is the normalized excitation force. $\theta_j$ is the electro-mechanical coupling coefficient and can be represented as:

$$\theta_j = -\frac{d_{31}Y_p b_p (h_c^2 - h_b^2)}{h_p}\phi_j'(x_2)\bigg|_{x_2=L_p} \tag{23}$$

where $Y_p$, $b_p$, $h_p$, and $e_{31}$ are the Young's modulus, width, thickness and piezoelectric constant of the piezoelectric layer. $h_b$ is the distance from the neutral axis of beam to the bottom surface of the piezoelectric layer and $h_c$ is the distance from the neutral axis of beam to the top surface of the piezoelectric layer.

The equivalent circuit of the system is depicted in Fig. 3. The equation of the circuit can be represented as:

$$C_p \dot{v}(t) + \frac{1}{R_l}v(t) + \sum_{j=1}^{2}\theta_j\dot{\eta}_j(t) = 0 \tag{24}$$

where $v$ is the output voltage, $R_l$ is the load resistance, and $C_p$ is capacitance of the piezoelectric layer.

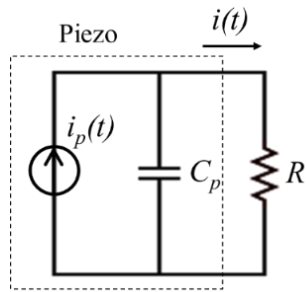

Fig. 3 The equivalent circuit of the PEH.

## 2.3. Model of the PEH with stoppers

The impulse force of stopper is introduced as an external force. Two typical scenarios are considered: collision between the main beam and the auxiliary beam, as shown in Fig. 1(c-d). The stopper has a stiffness much higher than that of the cut-out beam. The tip mass displacement of the main beam and auxiliary beam are limited by stopper #1 and stopper #2, respectively. In this model, the impact is considered as a completely inelastic collision and the stopper is treated as a 1-DOF system. Each impact force can then be represented as follows:

$$F_{imp_s} = \begin{cases} \dfrac{2\zeta_{st_s}\omega_{st_s}}{\left[\phi_{st_s}(L_{st_s})\right]^2}\left[\sum_{i=1}^{2}\phi_{s+2,j}(L_{s+2})\dot{\eta}_j(t)-\phi_{st_s}(L_{st_s})\dot{\eta}_{st_s}(t)\right] \\ +\dfrac{\omega_{st_s}^2}{\left[\phi_{st_s}(L_{st_s})\right]^2}\left[\sum_{i=1}^{2}\phi_{s+2,j}(L_{s+2})\eta_j(t)-\phi_{st_s}(L_{st_s})\eta_{st_s}(t)+d_s\right] & \text{if } y_{rel_s} \leq -d_s \\ 0 & \text{if } y_{rel_s} > -d_s \end{cases} \quad s=1,2 \tag{25}$$

where $F_{imp1}$ and $F_{imp2}$ represent the impact occurring when a stopper collides with the main and auxiliary beams, respectively. The subscript $s$ indicates the stopper number. $s$ equals 1 when the stopper is on the main beam and 2 when it's on the auxiliary beam. $d_1$ and $d_2$, as shown in Fig. 1(c-d), represent the distance between a stopper and the main and auxiliary beams. $y_{rel1}(t)$ and $y_{rel2}(t)$ represent the relative displacement between stopper and the main and auxiliary beam during vibration. By accounting for the impact force, the electromechanical coupling equation can be re-expressed as eqn. (26).

$$\begin{cases} \ddot{\eta}_j(t)+2\zeta_j\omega_j\dot{\eta}_j(t)+(\omega_j^2-Kg_j\cos\Omega t)\eta_j(t)+\left(Rs_j-\dfrac{1}{2}\Omega^2 Kn_j\right)\eta_j^3(t) \\ +Kn_j\left[\eta_j(t)\dot{\eta}_j^2(t)+\eta_j^2(t)\ddot{\eta}_j(t)\right]+\theta_j v(t)+F_{imp_s}\phi_{s+2,j}(L_{s+2})=F_j\sin\Omega t & j=1,2 \\ \ddot{\eta}_{st_s}(t)+2\zeta_{st_s}\omega_{st_s}\dot{\eta}_{st_s}(t)+\omega_{st_s}^2\eta_{st_s}(t)=F_{st_s}\sin\Omega t & s=1 \text{ or } 2 \\ Cp\dot{v}(t)+\dfrac{1}{R_l}v(t)+\sum_{j=1}^{2}\theta_j\dot{\eta}_j(t)=0 \end{cases} \tag{26}$$

## 2.4. Model of the PEH with magnetic force

In this section, the magnetic force is introduced to the PEH by applying it to either the tip of the main beam or the tip of the auxiliary beam, as depicted in Fig. 1(e-f). The magnet model is developed based on Akoun's model [37] for two cubic magnets and can be expressed as eqn.(27).

$$F_{mag}=-\nabla U=\dfrac{B_1 B_2}{4\pi\mu_0}\sum_{i=0}^{1}\sum_{j=0}^{1}\sum_{k=0}^{1}\sum_{l=0}^{1}\sum_{m=0}^{1}\sum_{n=0}^{1}(-1)^{i+j+k+l+m+n}\varphi_{x,y,z}(u,v,w,r) \tag{27}$$

where $B_1$ and $B_2$ are the magnetization intensity of the two magnet. $\mu_0$ is vacuum magnetic permeability.

$$\begin{cases} \varphi_x = \dfrac{1}{2}(v^2 - w^2)\ln(r-u) + uv\ln(r-v) + wv\tan^{-1}\left(\dfrac{uv}{wr}\right) + \dfrac{1}{2}ru \\ \varphi_y = \dfrac{1}{2}(u^2 - w^2)\ln(r-v) + uv\ln(r-u) + wu\tan^{-1}\left(\dfrac{uv}{wr}\right) + \dfrac{1}{2}rv \\ \varphi_z = -wu\ln(r-u) - wv\ln(r-v) + uv\tan^{-1}\left(\dfrac{uv}{wr}\right) - rw \end{cases} \quad (28)$$

$$\begin{cases} u = \alpha + (-1)^j \dfrac{a_1}{2} - (-1)^i \dfrac{a_2}{2} \\ v = \beta + (-1)^l \dfrac{b_1}{2} - (-1)^k \dfrac{b_2}{2} \\ w = \gamma + (-1)^n \dfrac{c_1}{2} - (-1)^m \dfrac{c_2}{2} \\ r = \sqrt{u^2 + v^2 + w^2} \end{cases} \quad (29)$$

where $\alpha$, $\beta$, and $\gamma$ are the X, Y, and Z component of the distance between the two magnets, respectively. $a_i$, $b_i$, and $c_i$ are the length, width, and height of the magnets, respectively.

In this model, $F_{mag}$ is in the Z direction as indicated in Fig. 1(b). By taking $F_{mag}$ into account, the electromechanical coupling equation of the PEH can be expressed as eqn.(30).

$$\begin{cases} \ddot{\eta}_j(t) + 2\zeta_j\omega_j\dot{\eta}_j(t) + (\omega_j^2 - Kg_j\cos\Omega t)\eta_j(t) + \left(Rs_j - \dfrac{1}{2}\Omega^2 Kn_j\right)\eta_j^3(t) \\ + Kn_j\left[\eta_j(t)\dot{\eta}_j^2(t) + \eta_j^2(t)\ddot{\eta}_j(t)\right] + \theta_j v(t) + F_{mag}\phi_{s+2,i}(L_{s+2}) = F_j\sin\Omega t \\ Cp\dot{v}(t) + \dfrac{1}{R_l}v(t) + \sum_{j=1}^{2}\theta_j\dot{\eta}_j(t) = 0 \end{cases} \quad j=1,2; s=1 \text{ or } 2 \quad (30)$$

## 3. Experiment

The experiment setup is shown in Fig. 4. The prototype (Fig. 4(a)) is fabricated from steel, with a PVDF film affixed to the auxiliary beam. The aluminum tip masses are fixed to the ends of the main beam and auxiliary beam with screws. Detailed parameters of the prototype are provided in Table 1. The experiment platform (Fig. 4(b)) features a controller that regulates rotational speed, conducting both up-sweep and down-sweep tests. The motor, controlled by speed signal inputs, provides rotational excitation to the PEH. The voltage of the PVDF is measured by an oscilloscope, enabling the recording of voltage response data throughout the tests.

Table 1 Parameters of the prototype.

| Parameters | Symbol | Value |
|---|---|---|
| Radius of rotation of the PEH | $R_l$ | 30 mm |
| Radius of rotation of the stopper | $R_{st1}/R_{st2}$ | 90/60 mm |
| Length of the main beam | L | 130 mm |
| Length of the 1st/2nd/3rd segment of the main beam | $L_1/L_2/L_3$ | 33 mm/34 mm/63 mm |
| Length of the auxiliary beam | $L_4$ | 41 mm |
| Length of the stopper beam | $L_{st1}/L_{st2}$ | 70/55 mm |
| Length of the PVDF patch | $L_e$ | 34 mm |
| Young's modulus of the substrate/stopper beam (steel) | $Y_s/Y_{st}$ | 193 GPa |
| Young's modulus of the PVDF patch | $Y_p$ | 45 GPa |
| Thickness of the substrate | $h_s$ | 0.3 mm |
| Thickness of the PVDF patch | $h_e$ | 0.4 mm |
| Thickness of the stopper | $h_{st1}/h_{st2}$ | 1 mm |
| Width of the main beam | $b_1$ | 20 mm |
| Width of the auxiliary beam | $b_2$ | 10 mm |
| Width of the PVDF patch | $b_e$ | 12 mm |
| Density of the substrate/stopper beam (steel) | $\rho_s/\rho_{st}$ | 7930 kg/m$^3$ |
| Density of the PVDF patch | $\rho_e$ | 1780 kg/m$^3$ |
| Tip mass of the $i^{th}$ segment | $Mt_i$ | 0 g/0 g/2.42 g/1.25 g |
| Capacitance (PVDF) | $C_p$ | 1.38 nF |
| Piezoelectric constant (PVDF) | $d_{31}$ | $-23\times10^{-12}$ C/N |
| Load resistance | $R_l$ | 1 MΩ |

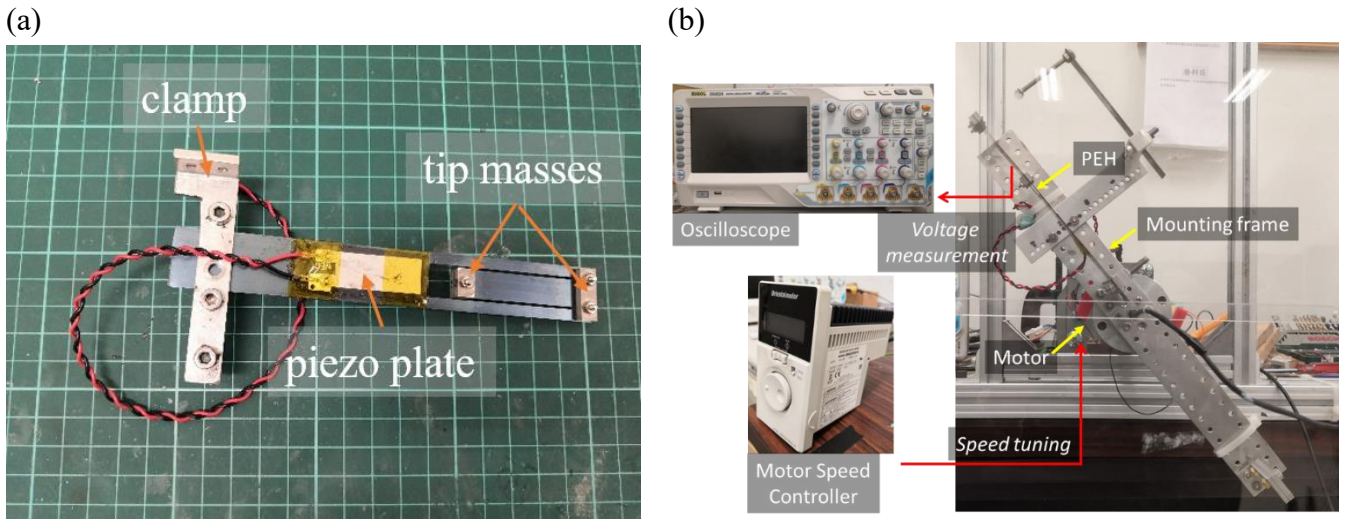

Fig. 4 Experiment setup (a) prototype (b) test platform.

## 4. Results

In this chapter, the proposed 2-DOF PEH and the version with stoppers or magnets are examined under rotational excitation to validate the theoretical models. The impact of three structural parameters of the PEH, tip mass of main beam ($M_{t3}$), tip mass of auxiliary beam ($M_{t4}$), and length of the main beam ($L$), on the performance of the PEH is examined by adjusting their values. The parameter settings for four different cases are listed in Table 2. In Section 4.2 and 4.3, the impact force and the magnetic force are respectively introduced to investigate their influence on the performance of the PEH.

Table 2 Parameter settings for four different cases

| Case | $M_{t3}$ | $M_{t4}$ | $L$ |
| --- | --- | --- | --- |
| Case A | 2.42 g | 1.25 g | 130 mm |
| Case B | **3.42 g** | 1.25 g | 130 mm |
| Case C | 2.42 g | **2.25 g** | 130 mm |
| Case D | 2.42 g | 1.25 g | **140 mm** |

**4.1. Rotational PEH with different parameter settings**

The PEH is first examined without impact force or magnetic force to assess the influence of centrifugal force on its responses under rotational excitation. Fig. 5 illustrates the natural frequencies and mode shapes under different excitation/driving frequencies for case A. In the Campbell diagram (Fig. 5(a)), which depicts the natural frequencies in relation to the driving frequency, initially, the first natural frequency increases while the second natural frequency decreases as the driving frequency increases. Once the driving frequency reaches approximately 14.7 Hz, the two mode curves reach their closest point. Beyond this point, the first resonant frequency begins to decrease while the second resonant frequency starts to increase as the driving frequency continues to rise. This phenomenon is known as "mode veering."

It is noted that the line labeled "perfect match" indicate when the natural frequency equals the driving frequency, signifying that the proposed rotational PEH operates at its resonance, delivering high power output. It can be seen in Fig. 5(a) that the first mode curve intersects the perfect match line at 13.2 Hz, while the second mode curve intersects the perfect match line at 14.9 Hz. This indicates that the PEH will resonant at 13.2 Hz and 14.9 Hz under rotational excitation. Fig. 5(b) illustrates the first two mode shapes at various driving frequencies. Notably, the two mode shapes can be distinguished by examining the presence of an inflection point on the main beam. The second mode exhibits an inflection point in its mode shape.

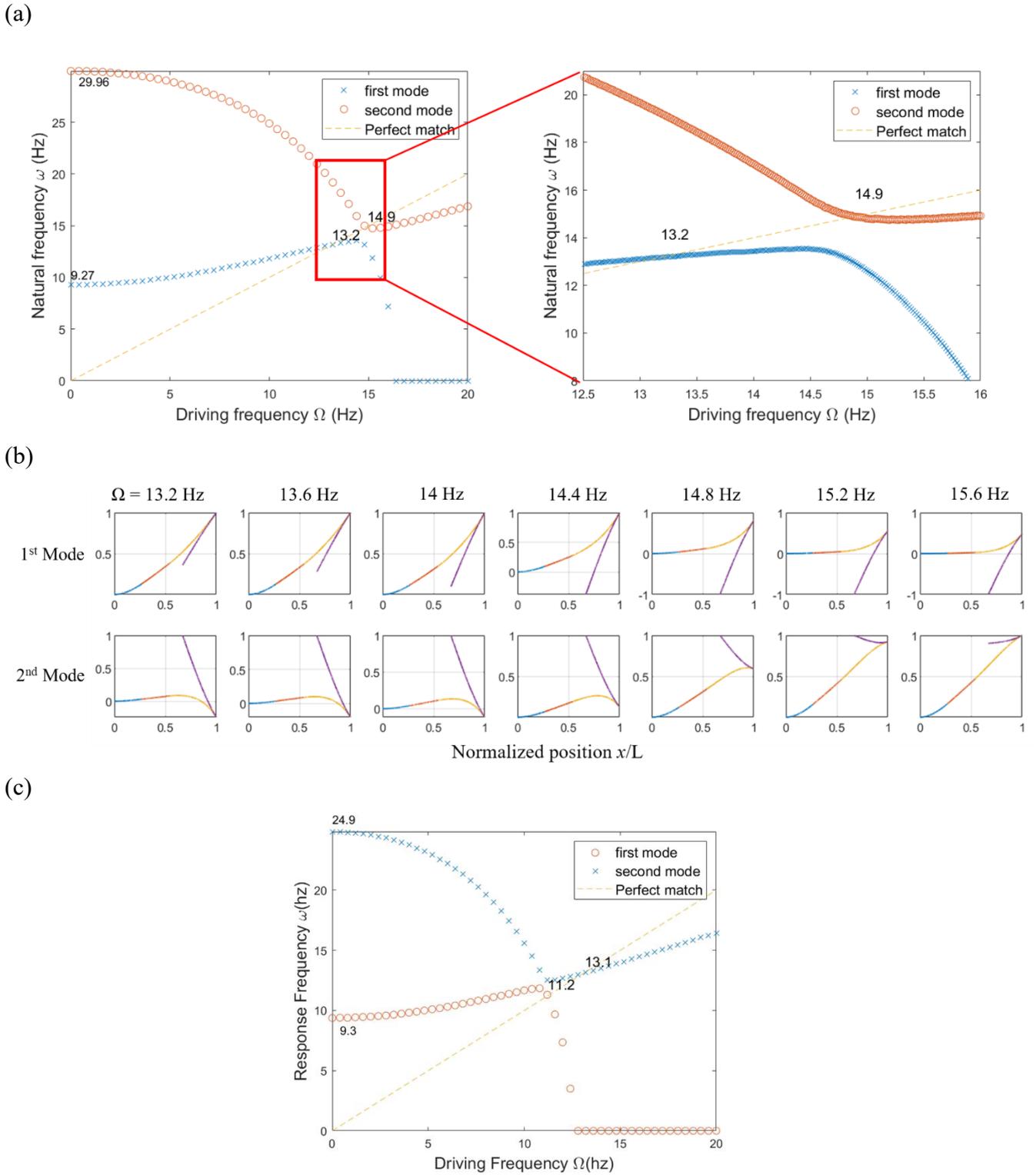

Fig. 5 Natural frequencies and mode shapes (a) the Campbell diagram and mode veering of case A (b) mode shapes at different driving frequency Ω of case A (c) the Campbell diagram and mode veering of case C.

Fig. 6 show the frequency responses of the four different cases listed in Table 2. Modifying the structural parameters can influence the voltage responses, which can be categorized into two types. The first type occurs when the frequency at which the perfect match line intersects with the second mode curve still allows the first mode to exists. Case A is an example of to the first type. In Fig. 5(a), the second mode curve intersects the perfect line at 14.9 Hz, where the first natural frequency is a real number. In this scenario, the voltage response shows two peaks with an anti-resonance in between. Cases A, C, and D belong to the first type, as shown in Fig. 6 (a, b, and d).

The second type occurs when the frequency at which the perfect match line intersects the second mode curve causes the first mode to vanish. Case C is an example of the second type. In Fig. 5(c), the second mode curve intersects the perfect line at 13.1 Hz, where the first resonant frequency becomes imaginary (indicated as zero in the plot). In this scenario, the voltage response exhibits only one peak but with a higher amplitude, as shown in Fig. 6 (c).

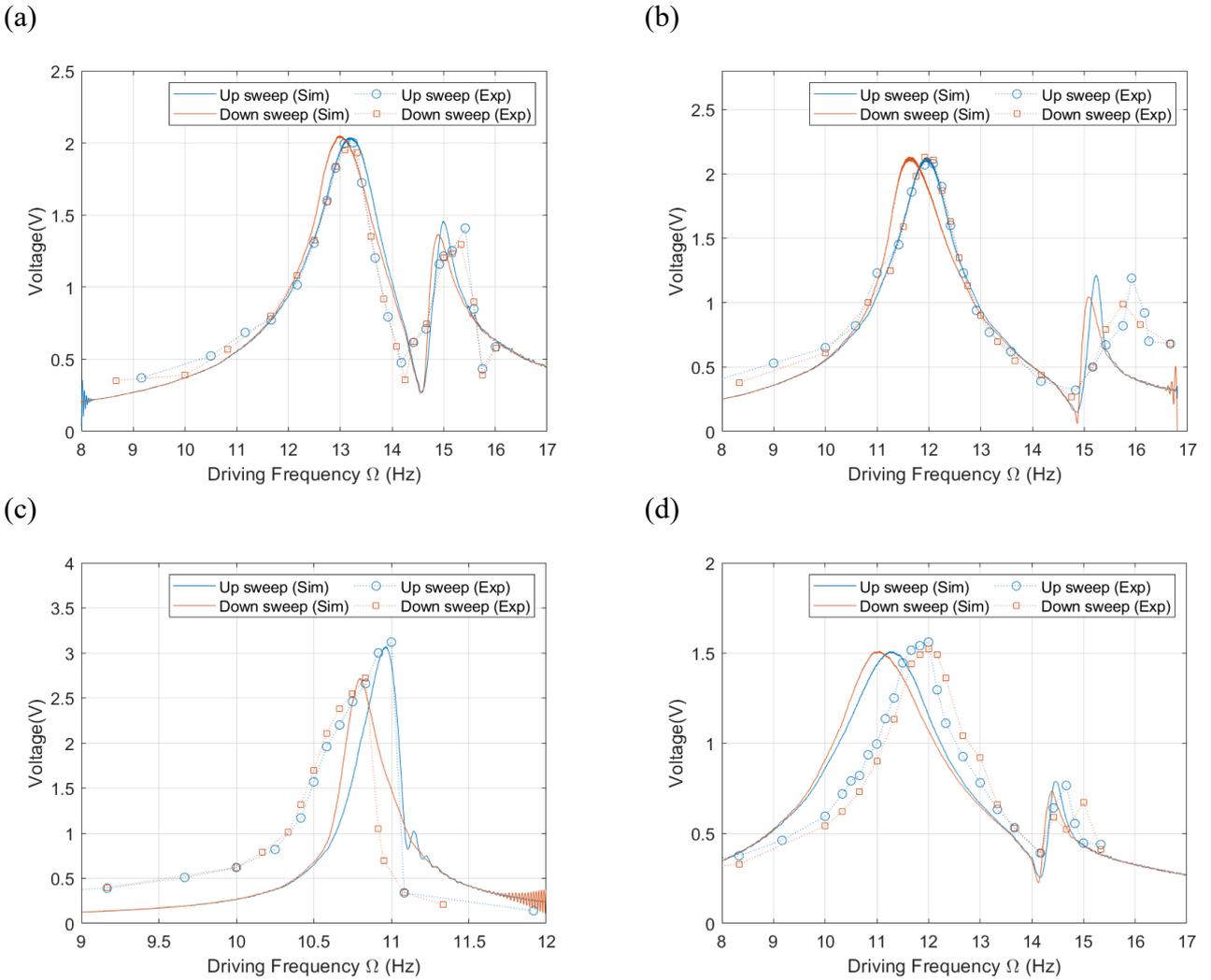

Fig. 6 Voltage responses of the PEH for four different cases (a) case A (b) case B (c) case C (d) case D

### 4.2. Rotational PEH with magnetic force

This section investigates the influence of magnetic force on the performance of the PEH under rotational excitation. The parameter settings for the PEH with magnetic force on the main beam is based on case B (Table 2). Fig. 7 depicts the voltage responses of the PEH with magnetic force on the main beam. The experimental results align well with the simulation results, validating the accuracy of the model. Both the experimental and simulated up- and down-sweep voltage responses exhibit the hardening effect, which enhances the up-sweep responses in terms of both bandwidth and amplitude. The hardening effect refers to a phenomenon in which the system stiffness increases with displacement. This behavior is often mathematically represented by cubic or higher-order stiffness terms. In this section, the observed hardening effect is attributed to the magnetic force. When the magnet gap $d_1$ is 39.5 mm, it can be observed that the peak frequency is 13.1 Hz with an amplitude of 1.95 V under the experimental up-sweep test. However, the peak frequency drops to 12.6 Hz with an amplitude of 1.58 V under the experimental down-sweep test. It is noted that the hardening effect is only observed around the first resonant frequency and not the second. This is due to the small tip displacement of the main beam in the second mode, resulting in a minimal influence of the magnetic force.

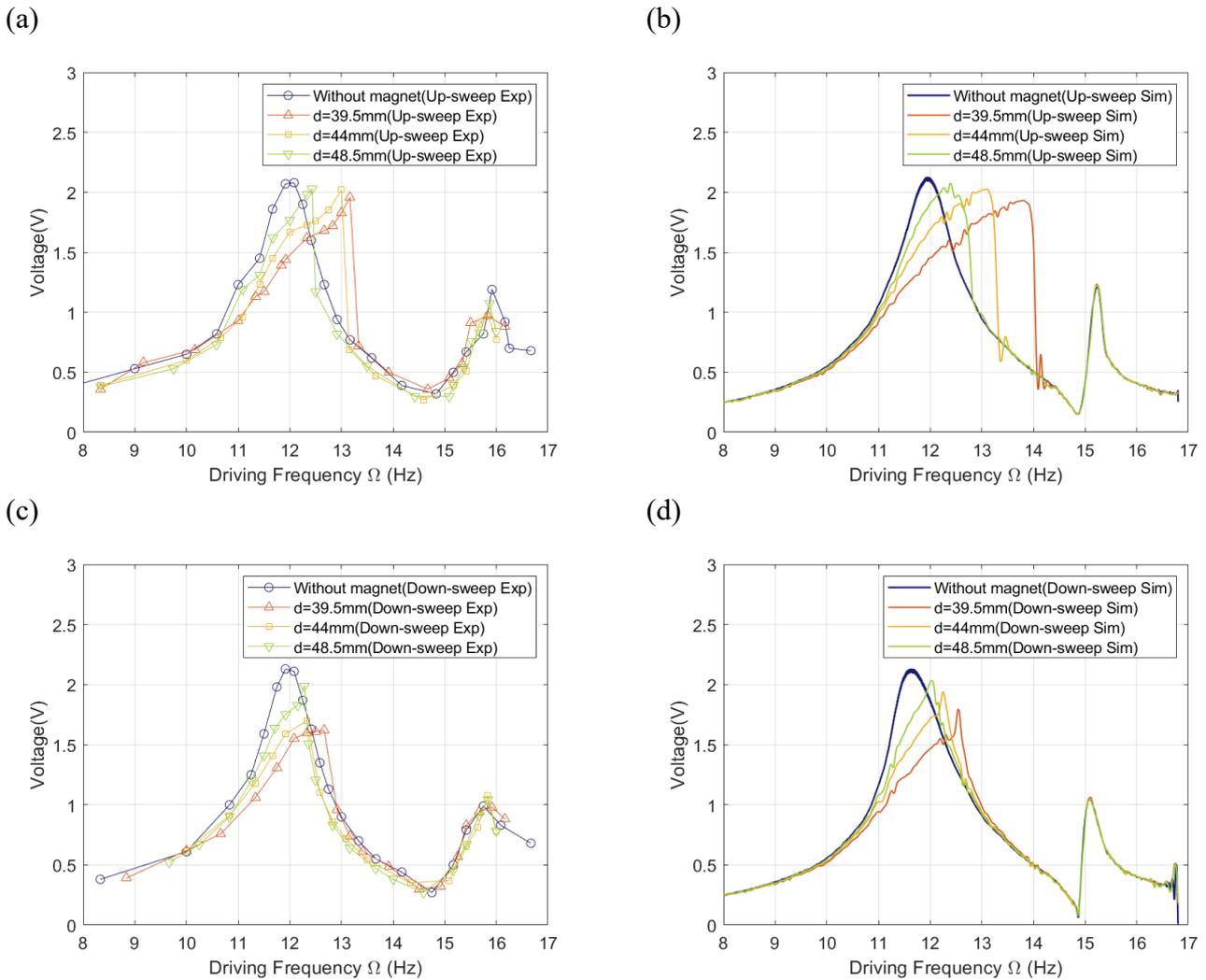

Fig. 7 Voltage responses of the PEH with magnets on the main beam (a) experiment, up-sweep (b) simulation, up-sweep (c) experiment, down-sweep (d) simulation, down-sweep.

The parameter settings for the PEH with magnetic force on the auxiliary beam is based on case C

(Table 2). Fig. 8 illustrates the voltage responses of the PEH. It is shown that the simulation and experimental results match well. Unlike the configuration with magnetic force on the main beam, the magnetic force on the auxiliary beam barely introduces the hardening effect to the system.

To compare the power output of the PEH with different configurations, this study introduces a parameter called the "power area" for performance evaluation. The power area is defined as:

$$\text{Power Area} = \int_{\omega_L}^{\omega_H} \frac{V^2}{R} d\omega \tag{31}$$

where $\omega_L$ and $\omega_H$ are the lower and upper bound frequencies to obtain the power area. The power generation with various stopper settings on the main beam and the auxiliary beam is presented in Table 5 and Table 6, respectively. It is noted that the "power efficiency" is calculated by dividing the power area of the setting with no stopper during an up sweep by the power area of the compared setting. The results of magnets on main beam indicate that the bandwidth of the proposed PEH can be broadened by 0.75 Hz at the first resonant frequency. As indicated in Table 3, all the configurations with magnets exhibit signficant drops in power efficiency. The magnetic force on auxiliary beam shows no benefit on the bandwidth. As can be seen in Table 4, the efficiency loss is about 34.5% during up sweeps and 9%-18% during down sweeps.

(a)
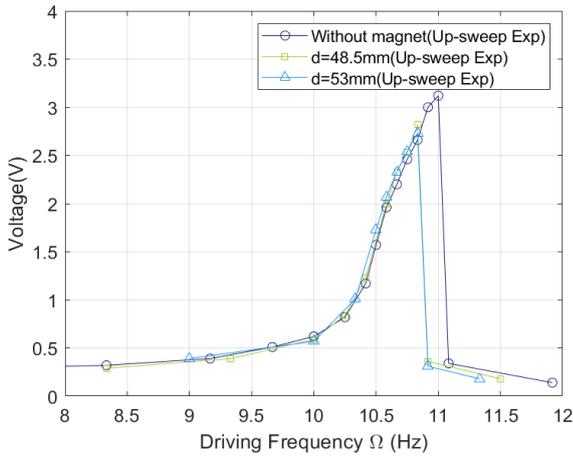

(b)
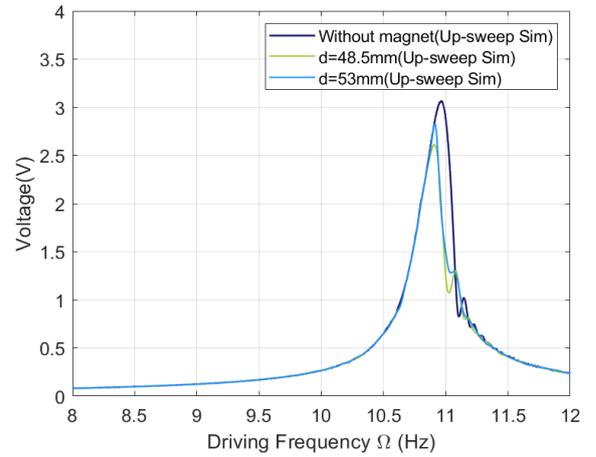

(c)
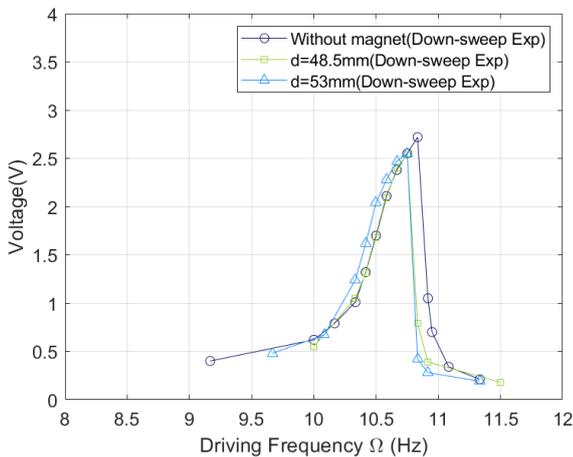

(d)
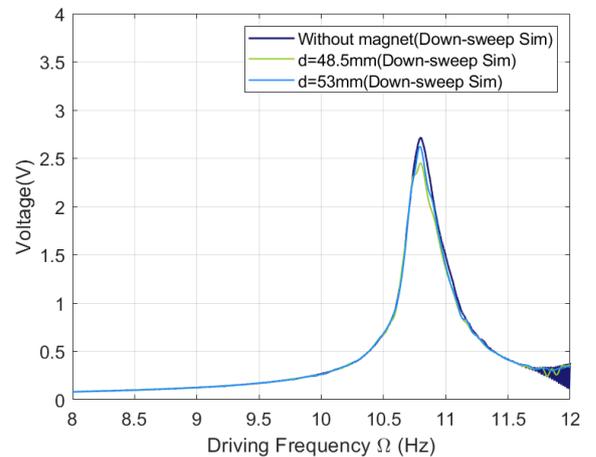

Fig. 8 Voltage responses of the PEH with magnets on the auxiliary beam (a) experiment, up-sweep (b) simulation, up-sweep (c) experiment, down-sweep (d) simulation, down-sweep.

Table 3 Comparison of power generation with various magnet settings on the main beam.

| Setting | Sweep | Power area (mW·Hz) ($\omega_L$ = 10.0 Hz, $\omega_H$ = 16.7 Hz) | Power efficiency | Bandwidth increase (Hz) 1st | 2nd |
|---|---|---|---|---|---|
| No magnet | up | 7.37×10⁻³ | 100% | - | - |
|  | down | 7.43×10⁻³ | 100.9% | - | - |
| $d_1$ = 48.5 mm | up | 5.66×10⁻³ | 76.8% | 0.75 | 0 |
|  | down | 5.56×10⁻³ | 75.4% | 0.17 | 0 |
| $d_1$ = 44 mm | up | 6.85×10⁻³ | 93.0% | 0.67 | 0 |
|  | down | 5.23×10⁻³ | 71.0% | 0 | 0 |
| $d_1$ = 39.5 mm | up | 6.85×10⁻³ | 92.9% | 0.18 | 0 |
|  | down | 5.43×10⁻³ | 73.6% | 0 | 0 |

Table 4 Comparison of power generation with various magnet settings on the auxiliary beam.

| Setting | Sweep | Power area (mW·Hz) ($\omega_L$ = 9.58 Hz, $\omega_H$ = 11.3 Hz) | Power efficiency | Bandwidth increase (Hz) 1st | 2nd |
|---|---|---|---|---|---|
| No magnet | up | 3.96×10⁻³ | 100% | - | - |
|  | down | 2.75×10⁻³ | 69.53% | - | - |
| $d_2$ = 48.5 mm | up | 2.58×10⁻³ | 65.24% | 0 | - |
|  | down | 2.06×10⁻³ | 52.11% | 0 | - |
| $d_2$ = 53 mm | up | 2.60×10⁻³ | 65.63% | 0 | - |
|  | down | 2.36×10⁻³ | 59.79% | 0 | - |

### 4.3. Rotational PEH with stoppers

In this section, the stopper is respectively added to the main beam and auxiliary beam to examine its influence on the performance. The parameter settings of the PEH with stopper are based on case A (Table 2). Fig. 9 shows the results when the stopper is on the main beam. It is noted that the first peak, related to the first resonant frequency, shifts to a higher frequency due to the hardening effect, similar to the configuration with magnetic force. The second peak, related to the second resonant frequency, remains nearly unchanged. This is because the deflection of main beam in the second resonant frequency is small, minimizing the influence of the stopper. When the gap $d_1$ reduces from 18.9 mm to 14.4 mm, the hardening effect shifts the peak frequency toward a higher frequency range but with reduced amplitude, as the stopper limits the displacement.

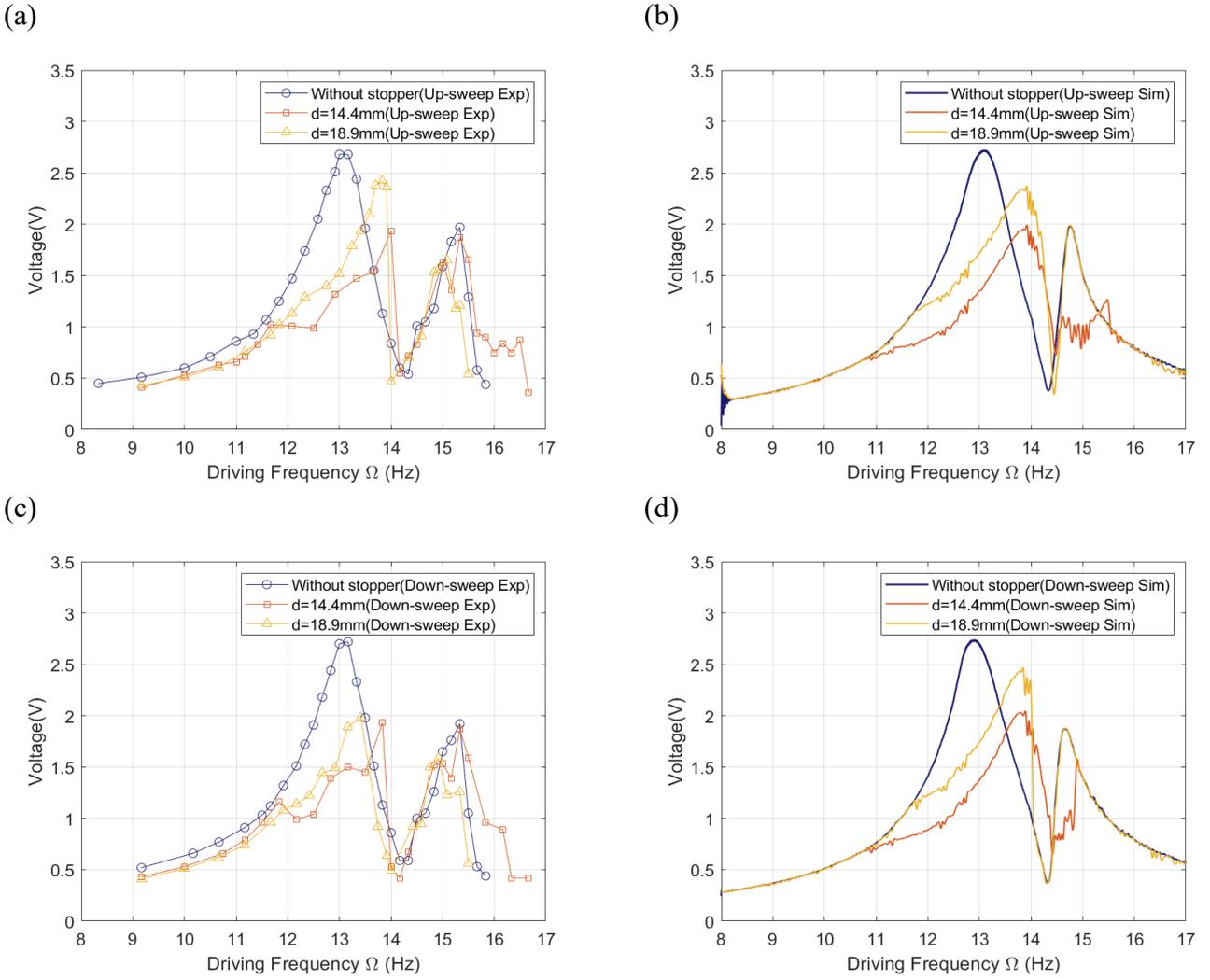

Fig. 9 Voltage responses of the PEH with a stopper on the main beam (a) experiment, up-sweep (b) simulation, up-sweep (c) experiment, down-sweep (d) simulation, down-sweep.

Fig. 10 depicts the PEH with a stopper on the auxiliary beam under rotational excitation. It can be seen that introducing a stopper on the auxiliary beam slightly decreases the amplitude of the first peak, but stronger effect is observed at the second peak. When $d_1$ is reduced to 15.4 mm, the hardening effect widens the bandwidth under up sweeps. However, the amplitude is reduced due to the displacement limit imposed by the stopper.

Table 5 and Table 6 present the power area and efficiency for the configurations with stoppers on the main and auxiliary beams, respectively. The results indicate that the proposed PEH can broaden the bandwidth by 0.42 Hz and 0.84 Hz at the first and second resonant frequencies, respectively, when a stopper is applied to the main beam. It is noted that the stopper causes approximately 30% power loss during up sweep and up to 34% during down sweep. For the configuration with a stopper on the auxiliary beam, the PEH broadens the bandwidth by 0.83 Hz at the second resonant frequency. Here, the stopper results in a power loss of 21% during the up sweep and 21%-22% during the down sweep.

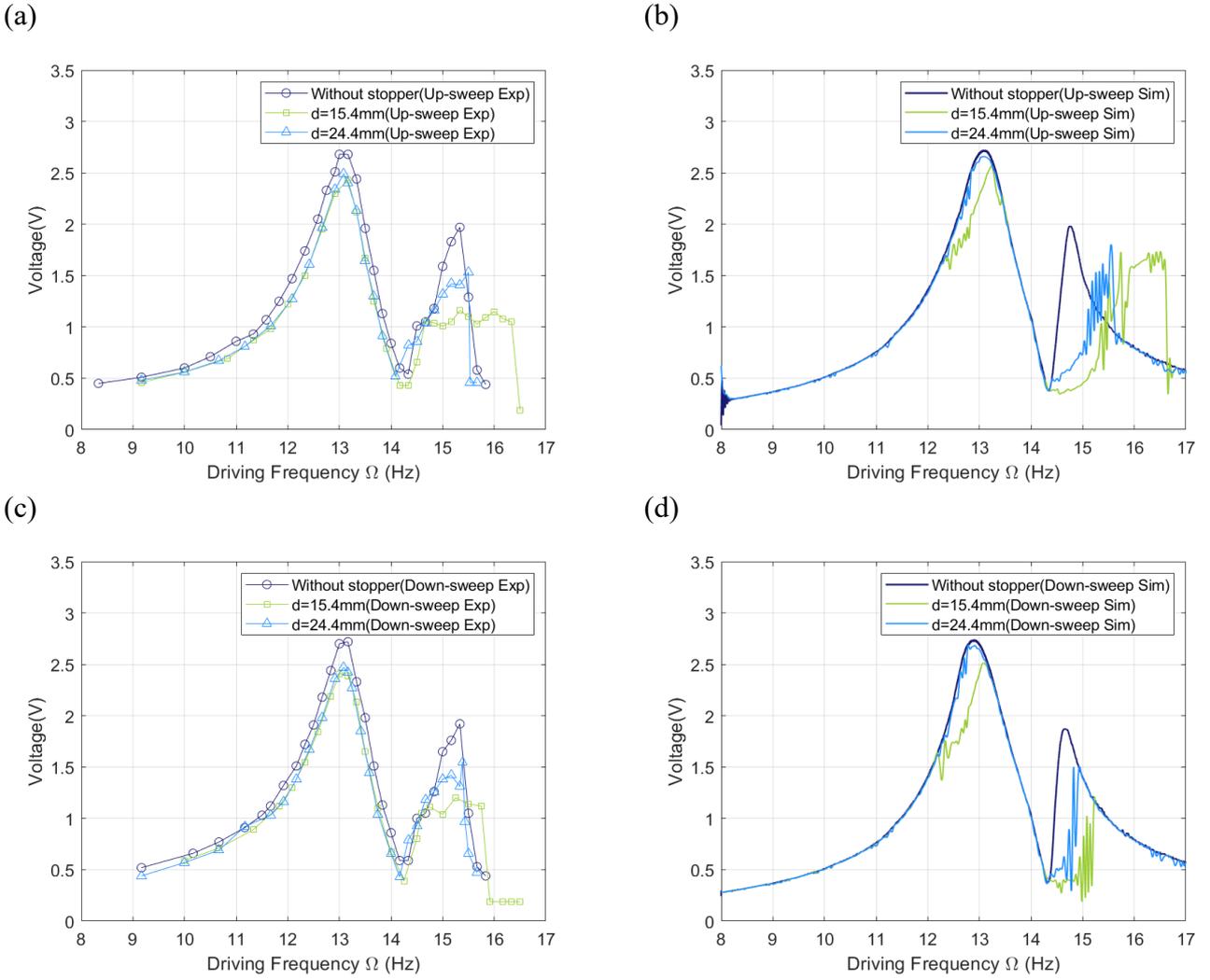

Fig. 10 Voltage responses of the PEH with a stopper on the auxiliary beam (a) experiment, up-sweep (b) simulation, up-sweep (c) experiment, down-sweep (d) simulation, down-sweep.

Table 5 Comparison of power generation with various stopper settings on the main beam.

| Setting | Sweep | Power area (mW·Hz) ($\omega_L$ = 11.2 Hz, $\omega_H$ = 16.7 Hz) | Power efficiency | Bandwidth increase (Hz) 1st | 2nd |
|---|---|---|---|---|---|
| No stopper | up | 12.5×10$^{-3}$ | 100% | - | - |
|  | down | 12.4×10$^{-3}$ | 98.94% | - | - |
| $d_1$ = 14.4 mm | up | 8.74×10$^{-3}$ | 69.68% | 0.33 | 0.83 |
|  | down | 8.61×10$^{-3}$ | 68.65% | 0.16 | 0.84 |
| $d_1$ = 18.9 mm | up | 8.91×10$^{-3}$ | 71.06% | 0.42 | 0 |
|  | down | 8.15×10$^{-3}$ | 64.99% | 0 | 0 |

Table 6 Comparison of power generation with various stopper settings on the auxiliary beam.

| Setting | Sweep | Power area (mW·Hz) ($\omega_L$ = 11.2 Hz, $\omega_H$ = 16.7 Hz) | Power efficiency | Bandwidth increase (Hz) 1st | 2nd |
|---|---|---|---|---|---|
| No stopper | up | 12.5×10$^{-3}$ | 100% | - | - |
|  | down | $12.4\times10^{-3}$ | 98.94% | - | - |
| $d_2 = 15.4$ mm | up | $9.92\times10^{-3}$ | 79.10% | 0 | 0.83 |
|  | down | $9.62\times10^{-3}$ | 76.73% | 0 | 0.25 |
| $d_2 = 24.4$ mm | up | $9.86\times10^{-3}$ | 78.58% | 0 | 0 |
|  | down | $9.79\times10^{-3}$ | 78.09% | 0 | 0 |

## 5. Conclusion

In this paper, a 2-DOF PEH subjected to rotational excitation is studied. The PEH exhibits two closely located resonant frequencies within the low-frequency range. The theoretical model, which includes geometric nonlinearity arising from the significant displacement of the beam due to rotational excitations, demonstrates simulation results that align well with experimental results. In the tests of the PEH without stoppers and magnetic force, mode veering, where the first two modes interchange, can be observed as the rotational speed of the excitation varies. The proposed PEH demonstrates the capability to have two closely located resonant frequencies for energy harvesting. By appropriately adjusting the parameters, the two peaks corresponding to the two resonant frequencies can merge into a single peak with high amplitude. Additionally, stoppers and magnetic force are introduced to the PEH separately to examine their impact on the performance of the PEH under rotational excitations. These two nonlinearities demonstrated their capability to broaden the bandwidth due to the hardening effect. However, the power output observed in experiments decreases after introducing these nonlinearities due to the associated damping and limited displacement. The experimental results indicate that the bandwidth of the proposed PEH can be broadened by at least 0.75 Hz at the first resonant frequency when magnets act on the main beam and by at least 0.84 Hz at the second resonant frequency when a stopper is applied to the main beam.

**CRediT authorship contribution statement**

**Xiang-Yu Li**: Data curation, Investigation, Software, Validation, Visualization, Writing - original draft.
**I-Chie Huang**: Methodology, Software.
**Wei-Jiun Su**: Conceptualization, Methodology, Funding acquisition, Project administration, Resources, Supervision, Validation, Visualization, Writing - review & editing.

**Acknowledgments**

The authors gratefully acknowledge the financial support of the National Science and Technology Council of Taiwan under grant MOST 110-2 21-E-002-141-MY3.

# Appendix

## A. Force equilibrium of an infinitesimal beam element

The force equilibrium of an infinitesimal beam element is shown in Fig. 11. The force equilibrium equation in the Z direction is expressed as:

$$\begin{cases} \dfrac{\partial}{\partial x_i}\left[F_i \cos(\alpha_i(x_i,t))\right] - \dfrac{\partial}{\partial x_i}\left[Q_i \sin(\alpha_i(x_i,t))\right] = -m_i\left[\dfrac{\partial^2 u_i(x_i,t)}{\partial t^2} - \Omega^2\left(u_i(x_i,t) - x_i - R_i\right)\right] + m_i g \cos\Omega t \quad i=1,2,3 \\ \dfrac{\partial}{\partial x_4}\left[F_4 \cos(\alpha_4(x_4,t))\right] - \dfrac{\partial}{\partial x_4}\left[Q_4 \sin(\alpha_4(x_4,t))\right] = -m_4\left[\dfrac{\partial^2 u_4(x_4,t)}{\partial t^2} - \Omega^2\left(u_4(x_4,t) - x_4 - R_4\right)\right] - m_4 g \cos\Omega t \end{cases}$$

(A1)

where $F$ is the force in the direction of $\hat{t}$, $Q$ is the shear force in the direction of $\hat{n}$, and $\alpha$ is angle of beam element during vibration. The force equilibrium equation in the X direction is expressed as:

$$\begin{cases} \dfrac{\partial}{\partial x_i}\left[F_i \sin(\alpha_i(x_i,t))\right] + \dfrac{\partial}{\partial x_i}\left[Q_i \cos(\alpha_i(x_i,t))\right] = -m_i\left[\dfrac{\partial^2 w_i(x_i,t)}{\partial t^2} - \Omega^2 w_i(x_i,t)\right] - m_i g \sin\Omega t \quad i=1,2,3 \\ \dfrac{\partial}{\partial x_4}\left[F_4 \sin(\alpha_4(x_4,t))\right] + \dfrac{\partial}{\partial x_4}\left[Q_4 \cos(\alpha_4(x_4,t))\right] = -m_4\left[\dfrac{\partial^2 w_4(x_1,t)}{\partial t^2} - \Omega^2 w_4(x_4,t)\right] + m_4 g \sin\Omega t \end{cases}$$

(A2)

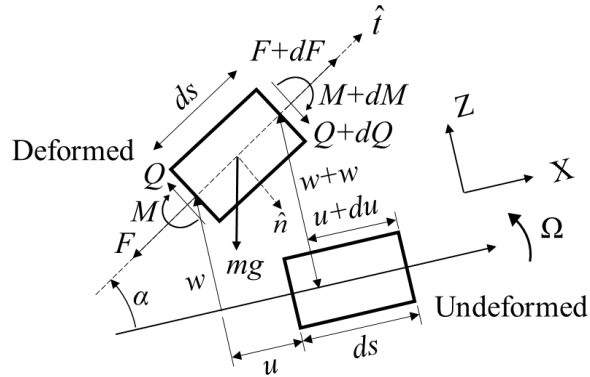

Fig. 11 Force equilibrium of an infinitesimal beam element

## B. Force equilibrium at the tip masses

Fig. 12 depicts the force equilibrium at the tip masses $M_{t3}$ and $M_{t4}$. The force equilibrium in the Z and X direction can be expressed as:

$$\begin{cases} -F_i\cos(\alpha_i(L_i,t))+Q_i\sin(\alpha_i(L_i,t))=-F_{i+1}\cos(\alpha_{i+1}(0,t))+Q_{i+1}\sin(\alpha_{i+1}(0,t)) \quad i=1,2 \\ \\ -F_3\cos(\alpha_3(L_3,t))+Q_3\sin(\alpha_3(L_3,t))-F_4\cos(\alpha_4(0,t))+Q_4\sin(\alpha_4(0,t))-Mt_3 g\cos\Omega t \\ =-Mt_3\left[\dfrac{\partial^2 u_3(L_3,t)}{\partial t^2}-\Omega^2\left(u_3(L_3,t)-L_3-R_3\right)\right] \\ \\ -F_4\cos(\alpha_4(L_4,t))+Q_4\sin(\alpha_4(L_4,t))+Mt_4 g\cos\Omega t=-Mt_4\left[\dfrac{\partial^2 u_4(L_4,t)}{\partial t^2}-\Omega^2\left(w_4(L_4,t)-L_4-R_4\right)\right] \end{cases}$$ (B1)

$$\begin{cases} -F_i\sin(\alpha_i(L_i,t))-Q_i\cos(\alpha_i(L_i,t))=-F_{i+1}\sin(\alpha_{i+1}(0,t))-Q_{i+1}\cos(\alpha_{i+1}(0,t)) \quad i=1,2 \\ \\ -F_3\sin(\alpha_3(L_3,t))-Q_3\cos(\alpha_3(L_3,t))-F_4\sin(\alpha_4(0,t))-Q_4\cos(\alpha_4(0,t))+Mt_3 g\sin\Omega t \\ =Mt_3\left[\dfrac{\partial^2 w_3(L_3,t)}{\partial t^2}-\Omega^2 w_3(L_3,t)\right] \\ \\ -F_4\sin(\alpha_4(L_2,t))-Q_4\cos(\alpha_4(L_4,t))-Mt_4 g\sin\Omega t=Mt_4\left[\dfrac{\partial^2 w_4(L_4,t)}{\partial t^2}-\Omega^2 w_4(L_4,t)\right] \end{cases}$$ (B2)

It is noted that $Mt_i g\sin\Omega t$ and $Mt_i g\cos\Omega t$ are the X and Z components of the gravitational force.

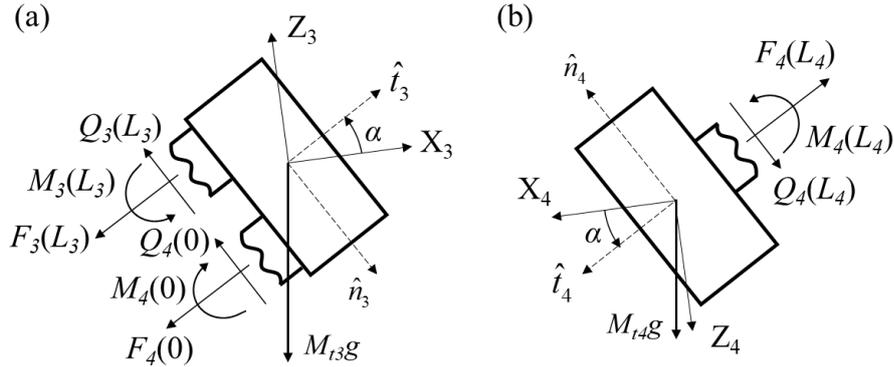

Fig. 12 Force equilibrium at (a) $Mt_3$ (b) $Mt_4$

## C. Terms of the undamped equation of motion of the PEH

$$Mb_{jk}=\sum_{i=1}^{4}\int_0^{L_i}\phi_{ik}(x_i)m_i\phi_{ij}(x_i)dx_i+\phi_{ik}(x_i)Mt_i\phi_{ij}(x_i)\Big|_{x_i=L_i}$$ (C1)

$$Kb_{jk}=\sum_{i=1}^{4}\int_0^{L_i}\dfrac{d^2\phi_{ik}(x_i)}{dx_i^2}YI_i\dfrac{d^2\phi_{ij}(x_i)}{dx_i^2}dx_i$$ (C2)

$$Ks_{jk}=\sum_{i=1}^{3}\int_0^{L_i}\dfrac{d\phi_{ik}(x_i)}{dx_i}\left[fc_i(x_i)+fc_{i+1}(0)\right]\dfrac{d\phi_{ij}(x_i)}{dx_i}dx_i+\int_0^{L_4}\dfrac{d\phi_{4k}(x_4)}{dx_4}fc_4(x_4)\Omega^2\dfrac{d\phi_{4j}(x_4)}{dx_4}dx_4$$ (C3)

$$Kg_{jk} = \sum_{i=1}^{3}\int_{0}^{L_i}\frac{d\phi_{ik}(x_i)}{dx_i}\left[fg_i(x_i) - fg_{i+1}(0)\right]\frac{d\phi_{ij}(x_i)}{dx_i}dx_i - \int_{0}^{L_4}\frac{d\phi_{4k}(x_4)}{dx_4}fg_4(x_4)\Omega^2\frac{d\phi_{4j}(x_4)}{dx_4}dx_4 \tag{C4}$$

$$Rs_{jk} = \sum_{i=1}^{4}\int_{0}^{L_i}\frac{\partial\phi_{ik}(x_i,t)}{\partial x_i}YI_i\frac{\partial\phi_{ij}(x_i,t)}{\partial x_i}\frac{\partial\left(\frac{\partial\phi_{ij}(x_i,t)}{\partial x_i}\frac{\partial^2\phi_{ij}(x_i,t)}{\partial x_i^2}\right)}{\partial x_i}dx_i \tag{C5}$$

$$Kn_{jk} = \sum_{i=1}^{3}\int_{0}^{L_i}\frac{d\phi_{ik}(x_i)}{dx_i}\left[\overline{fn_i}(x_i) + \overline{fn_{i+1}}(0)\right]\frac{d\phi_{ij}(x_i)}{dx_i}dx_i + \int_{0}^{L_4}\frac{d\phi_{4k}(x_4)}{dx_4}\overline{fn_4}(x_4)\Omega^2\frac{d\phi_{4j}(x_4)}{dx_4}dx_4 \tag{C6}$$

where $\overline{fn_i}(x_i)$ is the real part of $fn_i(x_i,t)$.

$$F_j = \sum_{i=1}^{3}\left[\int_{0}^{L_i}\phi_{ij}(x_i)gdx_i + \phi_{ij}(L_i)Mt_ig\right] - \int_{0}^{L_4}\phi_{4j}(x_4)gdx_4 - \phi_{4j}(x_4)Mt_4g\big|_{x_i=L_i} \tag{C7}$$